\documentstyle[aps,prl,multicol,epsf]{revtex}
\newcommand{\mb}[1]{ { \mbox{\boldmath{$#1$}}}  }

\begin{document}
\widetext
 
\title{Effect of disorder on superconductivity \\ 
       in the boson-fermion model}

\author{T. Doma\'nski$^{(a,b)}$ and K.I. Wysoki\'nski$^{(b)}$}
\address{$^{(a)}$
       Centre de Recherches sur les Tr\`es Basses Temp\'eratures 
       CNRS, 38-042 Grenoble Cedex 9, France
       \\ 
       $^{(b)}$
       Institute of Physics, Maria Curie Sk\l odowska
       University, 20-031 Lublin, Poland}

\date{\today}
\maketitle
\draft

\begin{abstract}
We study how a randomness of either boson or fermion site
energies affects the superconducting phase of the boson 
fermion model. We find that, contrary to what is expected 
for $s$-wave superconductors, the non-magnetic disorder 
is detrimental to the $s$-wave superconductivity. However, 
depending in which subsystem the disorder is located, we 
can observe different channels being affected. Weak disorder 
of the fermion subsystem is responsible mainly for
renormalization of the single particle density of states
while disorder in the boson subsystem directly leads
to fluctuation of the strength of the effective pairing 
between fermions.
\end{abstract}
  
\begin{multicols}{2}
\narrowtext

\section{Introduction}
The boson fermion (BF) model is an example of a
microscopic theory of nonconventional superconductivity.
It describes a mixture of itinerant electrons or holes 
(fermions) which interact via charge exchange with 
a system of immobile local pairs (hard-core bosons). 
Due to this interactions, bosons acquire finite mass 
and under proper circumstances might undergo Bose 
condensation transition while fermions simultaneously 
start to form a broken symmetry superconducting
phase. 

For the first time this model has been introduced 
{\em ad hoc} almost two decades ago \cite{Ranninger-85} 
to describe the electron system coupled to the lattice 
vibrations in a crossover regime, between the adiabatic 
and antiadiabatic limits. Later it has been formally 
derived from the Hamiltonian of wide band electrons 
hybridized to the strongly correlated narrow band 
electron system \cite{Robaszkiewicz-87}. Very recently 
\cite{Auerbach-02} the same effective BF model has been 
derived purely from the two dimensional Hubbard model 
in the strong interaction limit using the contractor 
renormalization method of Morningstar and Weinstein 
\cite{plaquette-96}.

Some authors have proposed it as a possible scenario for 
description of high temperature superconductivity (HTSC). 
The unconventional way of inducing the superconducting
phase in the BF model has been independently investigated 
in a number of papers \cite{Eliashberg-87,Friedberg-89,%
Ioffe-89,Micnas-90,Ranninger-95,Geshkenbein-97,Micnas-01}. 
Moreover, this model reveals also several unusual 
properties of the normal phase (for $T>T_{c}$) with 
an appearance of the pseudogap being the most transparent 
amongst them \cite{perturbative,DMFT,Domanski-01}. 
Apart of eventual relevance of this model to HTSC 
there are attempts to apply the same type of picture 
for a description of the magnetically trapped atoms
of alkali metals \cite{Holland-01}.

The important question which we want to address in this paper 
is: what is an influence of disorder on superconductivity 
of the BF model ? The conventional $s$-wave symmetry 
BCS-type superconductors are known to be rather weakly 
affected by paramagnetic impurities \cite{Anderson-59} 
- the fact which is known as "Anderson theorem". 
Nonmagnetic impurities have remarkable detrimental 
effect on superconductors with the anisotropic order 
parameters. Magnetic impurities lead to pair-breaking 
effects which result in a strong reduction of $T_{c}$ 
even in $s$-wave superconductors. Studying the effect 
of impurities on the superconductors has always been
an established tool for investigation of the internal 
structure of the Cooper pairs. 

Due to the nonconventional pairing mechanism (i.e.\ 
exchange of the hard-core bosons between fermion pairs) 
it is  of a fundamental importance to see how the nonmagnetic 
impurities (disorder) affect the isotropic superconducting 
phase of the BF model.  
Previously, such a study has been carried out by
Robaszkiewicz and Paw\l owski \cite{Robaszkiewicz-01}
who considered disorder only in the boson subsystem.
Using a method of configurational averaging for the free 
energy, authors have shown a strong detrimental 
effect of disorder on superconductivity. Apart of a
reduction of the transition temperature $T_{c}$ they 
have also reported a remarkable change of a relative 
ratio $\Delta(T=0)/k_{B}T{c}$. In this paper we analyze 
the effect of disorder present in both: fermion and boson 
subsystems using a different method of the coherent 
potential approximation.

\section{The model and approach}

\subsection{Hamiltonian of the disordered BF model}

We consider the following Hamiltonian of the disordered 
BF model
\begin{eqnarray}
H^{BF} & = & \sum_{i,j,\sigma} t_{ij} c_{i\sigma}^{\dagger} 
c_{j\sigma} + \sum_{i} \left( \varepsilon_{i} - \mu \right)
c_{i\sigma}^{\dagger} c_{i\sigma}
\nonumber \\
& + &  \sum_{i} \left( \Delta_{B} + E_{i} - 2\mu \right) 
b_{i}^{\dagger} b_{i} 
\nonumber \\ & + &
v \sum_{i} \left(  b_{i}^{\dagger} c_{i\downarrow}
c_{i\uparrow} +  b_{i} c_{i\uparrow}^{\dagger}
c_{i\downarrow}^{\dagger} \right) \;.
\label{BF}
\end{eqnarray}
We use the standard notations for annihilation (creation) 
operators of fermion $c_{i,\sigma}$ ($c_{i,\sigma}^{\dagger}$) 
with spin $\sigma$ and of the hard core boson $b_{i}$ 
($b_{i}^{\dagger}$) at site $i$. Fermions interact
with bosons via the charge exchange interaction 
$v$ which is assumed to be local. There are two ways 
in which disorder enters into the consideration. 
Either (a) fermions are affected by it and this is 
expressed by the random site energies $\varepsilon_{i}$, 
or (b) hard core bosons via their random site energies 
$E_{i}$.
 
To proceed, we apply first the mean field decoupling 
for the boson fermion interaction
\begin{eqnarray}
b_{i}^{\dagger} c_{i\downarrow}c_{i\uparrow} 
\simeq \left< b_{i} \right>^{*}
c_{ i\downarrow} c_{i\uparrow} + 
b_{i}^{\dagger} \left< c_{ i\downarrow} 
c_{i\uparrow} \right> 
\label{decoupling}
\end{eqnarray} 
which is justified until $v$ is small enough in 
comparison to the kinetic energy of fermions.
After decoupling (\ref{decoupling}) we have to deal 
with the effective Hamiltonian composed of the separate 
fermion and boson contributions $H \simeq H^{F}+H^{B}$ 
\cite{Robaszkiewicz-87,Ranninger-95}
\begin{eqnarray}
H^{F} & = &  \sum_{i,j,\sigma} \left[ t_{ij}
+\delta_{ij} \left( \varepsilon_{i} - \mu \right) \right]
c_{i\sigma}^{\dagger} c_{j\sigma} 
\nonumber \\ & + & 
\sum_{i} \left(  \rho_{i}^{*} c_{i\downarrow}
c_{i\uparrow} +  \rho_{i} c_{i\uparrow}^{\dagger}
c_{i\downarrow}^{\dagger} \right)
\label{H_F} \\
H^{B} & =  & \sum_{i}\left [ \left( \Delta_{B} + E_{i} 
-2\mu \right) b_{i}^{\dagger} b_{i} +
x_{i} \; b_{i}^{\dagger} + x_{i}^{*} b_{i} \right] \;,
\end{eqnarray}
where $x_{i}=v\left< c_{i\downarrow} c_{i\uparrow} 
\right>$ and $\rho_{i}=v\left< b_{i} \right>$.  
The site dependence of $\rho_i$ and $x_i$ indicates 
the disorder induced amplitude fluctuations of
the order parameters. 
%This dependence extends to all sites in the system 
%and not only to the specified site $i$.  

\subsection{Boson part}

For a given configuration of disorder we can exactly 
find the eigenvectors and eigenvalues corresponding 
to the lattice site $i$ using a suitable unitary 
transformation. Statistical expectation values of 
the number operator $b_{i}^{\dagger}b_{i}$ and the 
parameter $\rho_{i}$ are given by 
\cite{Robaszkiewicz-87,Ranninger-95}
\begin{eqnarray}
\left< b_{i}^{\dagger} b_{i}\right> & = & 
\frac{1}{2} - \frac{\Delta_{B}+E_{i}-2\mu}{4\gamma_{i}} 
\tanh{\left(\frac{\gamma_{i}}{k_{B}T}\right)}, 
\label{nB} \\
\rho_{i} & = & - \; \frac{vx_{i}}{2\gamma_{i}}\tanh{\left(
\frac{\gamma_{i}}{k_{B}T}\right)}
\label{rho}
\end{eqnarray}
where $\gamma_{i}=\frac{1}{2}\sqrt{(\Delta_{B}+E_{i}
-2\mu)^{2}+4|x_{i}|^{2}}$ 
and $k_{B}$ is the Boltzmann constant. Note, that 
the site dependent fermion order parameter $x_i$ 
enters the expression for the boson number operator
(\ref{nB}) and the parameter $\rho_{i}$ (\ref{rho}).
Disorder of any subsystem is thus automatically
transfered onto the other one.  

\subsection{Fermion part}
Analysis of the fermion part (\ref{H_F}) is more
cumbersome. To study it we use the Nambu 
representation $\Psi_{i}^{\dagger}
=(c_{i\uparrow}^{\dagger}, c_{i\downarrow})$, 
$\Psi_{i}=(\Psi_{i}^{\dagger})^{\dagger}$
and define the matrix Green's function 
${\mb G}(i,j;\omega)=\left<\left<\Psi_{i};
\Psi_{j}^{\dagger}\right>\right>_{\omega}$. 
Equation of motion for this function reads
\begin{eqnarray}
\sum_l&~&
\left[ \begin{array}{cc} 
( \omega-\varepsilon_l+\mu ) \delta_{il}-t_{il} 
& - \rho_{i}^{*}\delta_{il} \\
  - \rho_{i}\delta_{il} &
( \omega+\varepsilon_l-\mu )\delta_{il}+t_{il} 
\end{array}
\right] \nonumber \\ &\times&
{\mb G}(l,j; \omega)={\mb 1} \delta_{ij}.
\label{eq2}
\end{eqnarray}

Using the matrix Green's function ${\mb G}^0(i,j;\omega)$
of a clean system
\begin{equation}
\label{GF_clean}
\left[ {\mb G}^{0}({\bf k};\omega) \right] ^{-1} 
= \left(\begin{array}{cc} \omega - \varepsilon_{\bf k} 
+ \mu & -\rho^{*} \\
-\rho & \omega + \varepsilon_{\bf k} - \mu
\end{array} \right) ,
\end{equation} 
where $\rho=v\frac{1}{N}\sum_{i}\left< \rho_{i} \right>$
is a global order parameter (which plays a role of the 
effective gap in the superconducting fermion subsystem),
and defining the single site impurity potential $\mb V_l$ 
as
\begin{equation}
\label{eq23}
{\mb V}_l  = \left(\begin{array}{cc}
\varepsilon_l & -\rho_{l}^{*} \\
-\rho_{l} & -\varepsilon_l
\end{array} \right)
\end{equation}
one can write down the following Dyson equation for
the Green's function ${\mb G}(i,j;\omega)$ 
\begin{equation}
\label{Dyson}
{\mb G}(i,j;\omega) = {\mb G}^0(i,j;\omega) + \sum_l
{\mb G}^0(i,l;\omega) {\mb V}_l {\mb G}(l,j;\omega).
\end{equation}
This Green's function depends on the specific disorder 
configuration. In order to pass through one usually
averages it over the all possible configurations.

For carrying out the configurational averaging we use 
a method of the Coherent Potential Approximation (CPA). 
The main idea of CPA is to replace the random potential 
${\mb V}_{l}$ by some uniform coherent potential 
${\mb \Sigma}(\omega)$. Formally, the Green's function 
which satisfies (\ref{Dyson}) with ${\mb V}_{l}$ replaced 
by ${\mb \Sigma}(\omega)$ is then given (in the momentum 
coordinates) by
\begin{equation}
\label{GF_CPA}
\left[ {\mb G}^{CPA}({\bf k};\omega) \right] ^{-1} 
= \left[ {\mb G}^0({\bf k};\omega) \right] ^{-1} 
- {\mb \Sigma}(\omega) . 
\end{equation}

Configuration at site $i$ is defined by values 
of the random energies $\varepsilon_{i}$, $E_{i}$ 
- we shall symbolically denote it by $\alpha \equiv
\left\{ \varepsilon_{i},E_{i} \right\}$. Any of 
possible configurations $\alpha$ can occur with some
probability $P(\left\{ \varepsilon_{i},E_{i} \right\})
\equiv c^{(\alpha)}$, and of course these probabilities 
are normalized $\sum_{\alpha} c^{(\alpha)}=1$. 

A particle propagating through the medium characterized 
by the coherent potential ${\mb \Sigma(\omega)}$ is thus,
at site $i$, scattered with probability $c^{(\alpha)}$
by the potential ${\mb V}^{(\alpha)}_{i}-{\mb \Sigma}
(\omega)$. For a chosen configuration $\alpha$ of the 
site $i$ the conditionally averaged local Greens function 
is given by
\begin{equation}
\label{GF_conditional}
\left[ {\mb G}^{(\alpha)}(i,i;\omega)\right] ^{-1} 
= \left[ {\mb G}^{CPA}(i,i;\omega)\right] ^{-1} 
- \left[{\mb V}^{(\alpha)}_{i} -
{\mb \Sigma}(\omega) \right] .
\end{equation}
This Green's function ${\mb G}^{(\alpha)}(i,i;\omega)$
describes the system in which all sites, except one
indicated by $i$, are described by the coherent potential
${\mb \Sigma}(\omega)$. In CPA one requires that, 
the average of the local Green's function is the 
same as the Green's function of the averaged system.
This CPA condition is identical with the following 
equation \cite{Wysokinski_78}
\begin{equation}
\label{CPA_constraint}
\sum_{\alpha} c^{(\alpha)}
{\mb G}^{(\alpha)}(i,i;\omega) = 
{\mb G}^{CPA}(i,i;\omega) .
\end{equation}

Equations (\ref{GF_CPA}--\ref{CPA_constraint}) 
have to be solved selfconsistently to yield
the coherent potential ${\mb \Sigma}(\omega)$.
Physical quantities such as fermion 
concentration $n^{F} \equiv \frac{1}{N} 
\sum_{i,\sigma} \left< c_{i\sigma}^{\dagger} 
c_{i\sigma} \right>$ and the superconducting
order parameter $x \equiv \frac{1}{N} \sum_{i}
x_{i} $ are to be calculated from
\begin{eqnarray}
n^{F} = - \; \frac{2}{\pi N} \int_{-\infty}^{\infty}
\frac{d\omega}{e^{\beta\omega}+1}
{\rm Im}\left\{ {\mb G}_{11}^{CPA}(i,i;
\omega + i0^{+}) \right\}
\label{nF}
\\
x = -v \; \frac{1}{\pi N} \int_{-\infty}^{\infty}
\frac{d\omega}{e^{\beta\omega}+1}
{\rm Im}\left\{ {\mb G}_{21}^{CPA}(i,i;
\omega + i0^{+}) \right\}
\label{x}
\end{eqnarray}
where $\beta=1/k_{B}T$.

In the following section we discuss the 
changes of the superconducting transition 
temperature $T_c$ caused by disorder.

\section{Disorder in fermion subsystem}

It is instructive to investigate the disorder
separately for fermion and boson subsystems. Let us 
start with fermion disorder $\varepsilon_{i}$. 
We set $E_{i}=0$ for all lattice sites. 
For the random fermion energies we choose
$\varepsilon_{i}=\varepsilon_{0}$ with probability
$c$ and $\varepsilon_{i}=0$ with probability $1-c$. 
It is a bimodal type disorder 
\begin{eqnarray}
P \left( \left\{ \varepsilon_{i} \right\} \right) 
= c \delta( \varepsilon_{i}-\varepsilon_{0} ) + 
\left( 1 - c \right) \delta( \varepsilon_{i}-0).
\end{eqnarray}  

Here we shall be mainly interested in the
superconducting transition temperature $T_c$. 
In this limit \cite{Wysokinski} the diagonal 
disorder affects mainly a diagonal part of 
the matrix Green's function ${\mb G}$. In fact, 
even for no disorder acting directly in bosonic 
subsystem the boson order parameter $\rho$ in 
equation (\ref{rho}) does depend on the site index 
via fermion order parameter $x_i$. However, we 
expect this {\em induced} disorder to be weak 
and neglect it. This allows us to show how disorder 
in fermionic subsystem only, affects $T_c$.

The off-diagonal elements of the coherent potential 
vanish. Due to the general symmetry $\Sigma_{22}(i\omega)
=-\Sigma_{11}(-i\omega)$ \cite{Wysokinski} we can
simplify the self-energy matrix to
\begin{eqnarray}
{\mb \Sigma}(i\omega) = \left(\begin{array}{cc}
\Sigma_{11}(i\omega) & 0 \\
0 & -\Sigma_{11}(-i\omega)
\end{array} \right).
\end{eqnarray}
$\Sigma_{11}(\omega)$ can be found from 
the CPA equation (\ref{CPA_constraint}) which, for 
a normal phase, takes a well known form 
\cite{Wysokinski_78}
\begin{eqnarray}
\label{CPA_standard}
\frac{1-c}{\left[ \Sigma_{11}(\omega)\right]^{-1}+F(\omega)}
+\frac{c}{\left[\Sigma_{11}(\omega)-\varepsilon_{0}
\right]^{-1}+F(\omega)} = 0
\end{eqnarray}
with $F(\omega)=\frac{1}{N}\sum_{\bf k} 
{\mb G}^{CPA}_{11}({\bf k},\omega)$. Equation 
(\ref{CPA_standard}) should be solved 
subject to a given dispersion relation
$\varepsilon_{\bf k}$ and parameters $c$,
$\varepsilon_{0}$.

Finally having calculated $\Sigma_{11}(\omega)$, 
we can find $n^{F}$ and $x$ (\ref{nF},\ref{x}) 
as well as $n^{B}$, $\rho$ (\ref{nB},\ref{rho}).
In particular, the critical temperature 
$T_{c}=\left( k_{B}\beta_{c}\right)^{-1}$ 
is given via
\begin{eqnarray}
1=v^{2} \; \frac{ {\rm tanh} \left[ \beta_{c} 
\left( \Delta_{B}-2\mu \right)/2 \right]}
{\Delta_{B}-2\mu}  \sum_{\bf k} 
\int_{-\infty}^{\infty} d\omega_{1}
\int_{-\infty}^{\infty} d\omega_{2}
\nonumber \\ \times
A({\bf k},\omega_{1}) A({\bf k},\omega_{2}) 
\frac{{\rm tanh} \left[ \beta_{c}\omega_{1}/2 \right]
+ {\rm tanh} \left[ \beta_{c}\omega_{2}/2 \right] }
{2 \left( \omega_{1} + \omega_{2} \right) }
\label{Tc19}
\end{eqnarray}
where $A({\bf k},\omega)=(-1/\pi) {\rm Im} \left\{ 
{\mb G}^{CPA}_{11}({\bf k},\omega + i0^{+}) \right\}$ denotes 
the  spectral function of the normal phase.

\begin{figure}
\centerline{\epsfxsize=6cm \epsfbox{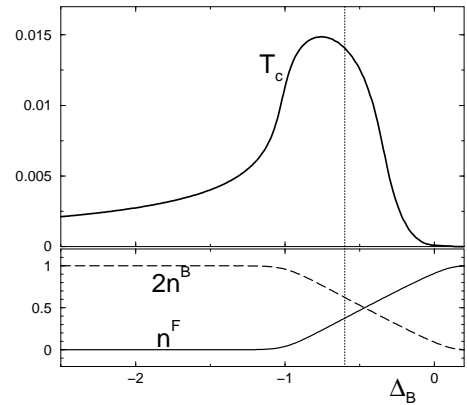}}
\vspace{3mm}
\caption{Variation of $T_{c}$ with respect to 
boson energy $\Delta_{B}$ for a clean system 
with $n_{tot}=1$. Bottom panel illustrates the
concentrations of fermions ($n^{F}$) and bosons 
($n^{B}$) at $T=T_{c}$. Note the three distinct regimes
of: predominantly local pairs $2n^{B} \sim n_{tot}$,
coexisting pairs and fermions $n^{F} \sim 2n^{B}$,
and predominantly fermions $n^{F} \sim n_{tot}$
(so called BCS limit).}
\label{fig0}
\end{figure}

We choose for our study a case of weak boson
fermion interaction $v=0.1$ (in units of the initial
fermion bandwidth) and total concentration of charge
carriers  $n_{tot} \equiv 2n_{B}+n_{F}=1$.
Figure (\ref{fig0}) shows how $T_{c}$ of a clean
system depends on position of the boson level.
There are three distinguishable regimes 
\cite{Robaszkiewicz-87,Micnas-90} of relative 
occupancy by bosons and fermions. Superconducting
correlations are of course most visible when
chemical potential is close to $\Delta_{B}/2$. 
We choose the value $\Delta_{B}/2=-0.3$ to be 
close to optimal value of transition temperature 
and to have comparable amount of fermions and bosons. 
  For  computations we use 
the 2D square lattice dispersion - the van Hove 
singularity is safely distant from the Fermi
energy for the above parameters.

%%%%%%%%%%%%%%%%%%%   Fig 1  %%%%%%%%%%%%%%%%%%%%%%%%%%
%%%%%%%%%%%%%%%%%%%%%%%%%%%%%%%%%%%%%%%%%%%%%%%%%%%%%%%

\begin{figure}
\centerline{\epsfxsize=6cm \epsfbox{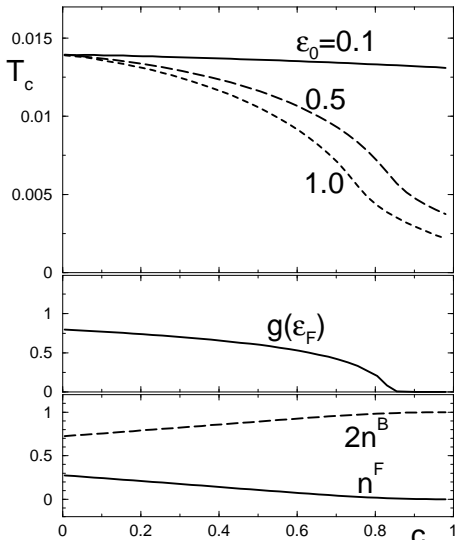}}
\caption{Transition temperature $T_{c}$ as a function 
of concentration $c$ of scattering centers with various 
positive values of $\varepsilon_{0}$ (top panel).
Density of states at the Fermi energy $g(\varepsilon_{F})$
(middle panel) and relative occupations by bosons and
fermions (bottom panel) for $\varepsilon_{0}=0.5$ at
$T=T_{c}$.}
\label{fig1}
\end{figure}

In figure \ref{fig1} we plot the transition temperature 
$T_{c}$, calculated from equation (\ref{Tc19})
 against concentration $c$ for several values of 
$\varepsilon_{0}$. With an increase of concentration $c$ 
of scattering centers we notice a gradual reduction of 
the critical temperature. This tendency can be understood 
by looking at the behavior of the fermion density of 
states at the Fermi energy $g(\varepsilon_{F})$. Disorder 
is responsible for renormalization of the low energy sector 
and these low energy states are involved in forming the
superconducting type correlations. As shown in the bottom 
panel there is additional effect coming from the rearrangement 
of occupations $n^{F}$ and $n^{B}$. With an increasing
concentration $c$ the fermion band is shifted toward higher
energies and the system is then mainly occupied by 
bosons (so called, the local pair LP limit).

%%%%%%%%%%%%%%%%%%%   Fig 2   %%%%%%%%%%%%%%%%%%%%%%%%%%%%%%
%%%%%%%%%%%%%%%%%%%%%%%%%%%%%%%%%%%%%%%%%%%%%%%%%%%%%%%%%%%

For negative values of $\varepsilon_{0}$ the disorder
shows stronger influence on $T_c$. On one hand we
have again a direct effect of the renormalized density
of states (see $g(\varepsilon_{F})$ in the middle panel
of figure 3). On the other hand, with an increase of
$c$ for any negative value of $\varepsilon_{0}$
the fermion band and the position of the chemical
potential drift to-wards lower energies. As its
consequence the number of fermions increases and
the number of bosons decreases. Effectively we thus
approach the BCS limit where transition temperature
diminishes very fast if $\Delta_{B}/2$ goes above $\mu$
(check for example the curves for $\varepsilon_{0}=-0.4$
and $-0.5$). The strong disorder in fermion subsystem
makes the pairing mechanism almost ineffective at all.

%%%%%%%%%%%%%%%%%%%   Fig 3   %%%%%%%%%%%%%%%%%%%%%%%%%%%%%%
\begin{figure}
\centerline{\epsfxsize=6cm \epsfbox{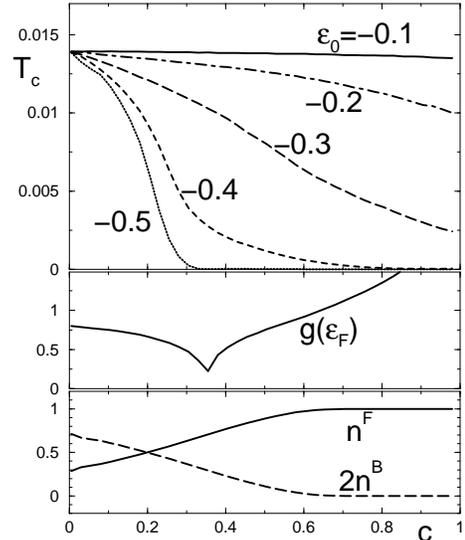}}
\caption{The same as in figure (\ref{fig1}) except
that for negative values of $\varepsilon_{0}$
(top panel). The middle and bottom panels correspond
to $\varepsilon_{0}=-0.5$.}
\label{fig2}
\end{figure}
%%%%%%%%%%%%%%%%%%%%%%%%%%%%%%%%%%%%%%%%%%%%%%%%%%%%%%%%%%%

\begin{figure}
\centerline{\epsfxsize=6cm \epsfbox{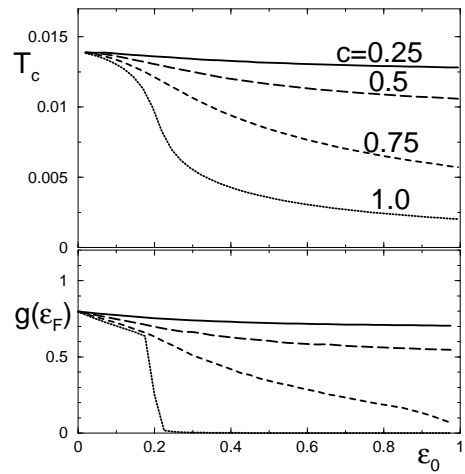}}
\vspace{3mm}
\caption{
Transition temperature $T_{c}$ as a function of 
the energy $\varepsilon_{0}$ of the scattering 
centers whose concentration is $c$ (top). Density
of states $g(\varepsilon_{F})$ for each of the
concentrations $c$ (bottom panel) For $c=1$
and for $\varepsilon_{0} \geq 0.2$ Fermi energy
goes below the fermions band, system is then
strictly in the LP limit of the BF model.} 
\label{fig3}
\end{figure}

In figure \ref{fig3} we plot $T_{c}$ versus (positive) 
$\varepsilon_{0}$  for several concentrations $c$. Again, 
$T_{c}$ roughly follows variation of the density of states 
$g(\epsilon_{F})$ which is shown in the bottom panel.
As discussed above for large values of concentration $c$
and large positive $\varepsilon_{0}$ the system is mainly
filled by bosons (the LP limit) so there is some finite
$T_{c}$ even when $\mu = \Delta_{B}/2$ is far below the
fermion band, this is an artifact of the mean field 
approximation \cite{Robaszkiewicz-87,Micnas-90}.

%%%%%%%%%%%%%%%%%%%   Fig 4   %%%%%%%%%%%%%%%%%%%%%%%
%%%%%%%%%%%%%%%%%%%%%%%%%%%%%%%%%%%%%%%%%%%%%%%%%%%%%

Behavior of $T_{c}$ with respect to negative values
of $\varepsilon_{0}$ can be easily deduced from the
figure 3 so we skip this illustration. 

In summary we notice that change of the transition 
temperature $T_{c}$ caused by weak disorder in fermion 
system is controlled mainly by modification of the 
low lying energy states. This is in accord with 
the Anderson theorem for spin singlet $s$-wave 
superconductors. However, additional influence 
comes from redistribution of particle spectrum 
and their relative occupancy and such effects are 
dominant for large values of impurity concentration 
$c$ and for their large scattering strength 
$|\varepsilon_{0}|$. In this limit the boson - 
fermion exchange becomes ineffective. 
%in driving the superconducting instability.

\section{Disorder in boson subsystem}

Now we turn attention to a case when boson energies 
are random $E_{i} \neq 0$ and, for simplicity, 
assume no fermion disorder i.e.\ $\varepsilon_{i}=0$ 
for all the lattice sites. The scattering potential 
(\ref{eq23}) reduces then to
\begin{equation}
\label{random_BFM}
{\mb V}_l  = \left(\begin{array}{cc} 0 & - f_{l} 
\left< c_{l\uparrow}^{\dagger}c_{l\downarrow}^{\dagger} \right> \\
- f_{l} \left< c_{l\downarrow}c_{l\uparrow} \right> & 0
\end{array} \right),
\end{equation}
with
\begin{equation}
\label{f_l}
f_{l} = v^{2} \; \frac{{\rm tanh}\left[ \beta \gamma_{l}
\right] }{2\gamma_{l}}
\end{equation}
and
\begin{equation}
\gamma_{l}=
\sqrt{ \left( \frac{\Delta_{B}+E_l}{2}
-\mu\right)^{2}+|v\left< c_{l\downarrow}c_{l\uparrow} \right>
|^{2}} \;.
\end{equation}
It means that the fluctuating boson energy level $E_l$ induces 
fluctuations of the pairing strengths $f_{l}$ in the fermion 
subsystem. To some extent, this situation reminds the 
negative $U$ Hubbard model \cite{Litak_00} for which 
the random local attraction $U_{l} < 0$ leads to 
the following scattering matrix
\begin{equation}
\label{random_negU}
{\mb V}^{(Hub)}_l  = \left(\begin{array}{cc}
U_{l} \left< n_{l} \right> /2 & U_{l} \left< 
c_{l\uparrow}^{\dagger} c_{l\downarrow}^{\dagger} \right> \\ 
U_{l} \left< c_{l\downarrow} c_{l\uparrow} \right>  & 
U_{l} \left< n_{l} \right> /2
\end{array}
\right).
\end{equation}
We see that in our case the role of a random 
pairing potential $U_{l}$ is played by $-f_{l}$ 
given in equation (\ref{f_l}).

There are two extreme limits, as far as the effectiveness of 
the random boson energy $\Delta_{B}+E_{l}$ is concerned 
\begin{itemize}
\item{for small (on the scale of fermion-boson interaction $v$)
fluctuations of $E_{l}$, effect of
   the disorder becomes negligible unless the chemical
   potential is pinned to the boson level $\mu =
   \Delta_{B}/2$, when the amplitude of the pairing 
   potential is controlled by 
   $f_{l} \sim v^{2} {\rm tanh} \left[ \beta x_{l}
   \right]/2x_{l}$ 
   and is usually uniform except at very low temperatures
   $\beta \rightarrow \infty$ when $f_{l} \sim v^2/x_l$}
\item{for large fluctuations of $E_{l}$ one obtains 
   $f_{l} \sim v^{2} {\rm tanh}\left[ \frac{\beta}{2}
   \left( \Delta_{B} + E_{l} -2\mu \right) \right] 
   / \left( \Delta_{B} + E_{l} -2\mu \right)$.}
\end{itemize}

\begin{figure}
\centerline{\epsfxsize=8cm \epsfbox{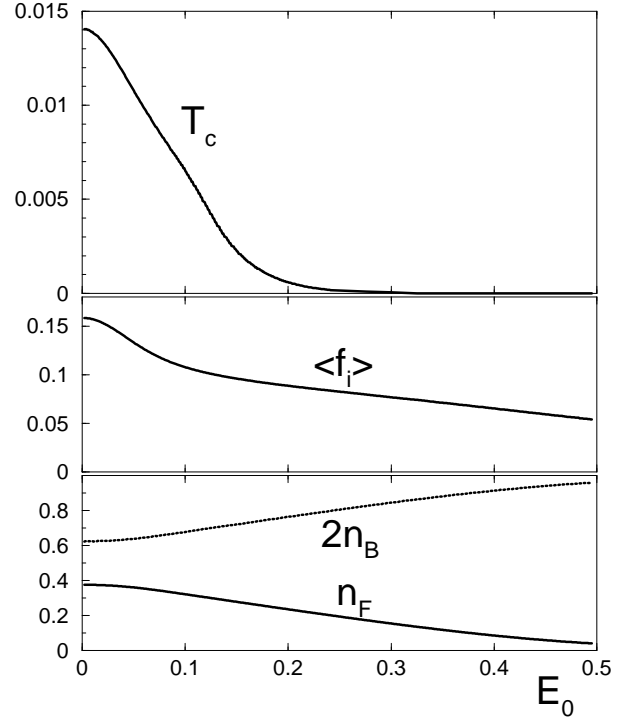}}
\vspace{3mm}
\caption{Transition temperature $T_{c}$ (top), 
the averaged pairing potential 
$<f_{l}> = \sum_{ \left\{ E_{l} \right\}}
P(\left\{ E_{l} \right\} ) f_{l}$ 
(middle), together with the occupation of fermions 
$n_{F}$ and bosons $n_{B}$ at $T=T_{c}$ (bottom).} 
\label{fig4}
\end{figure}

To analyze effects of the disorder in boson subsystem
we use a two pole distribution $P(\left\{ E_{l}
\right\} )=\frac{1}{2} \left[ \delta(E_{l}-E_{0}) +
\delta(E_{l}+E_{0}) \right]$. The boson energy is
$\Delta_{B} \pm E_{0}$ with an equal probability $0.5$.
Figure (\ref{fig4}) shows critical temperature $T_{c}$ ,
calculated from equation (\ref{x}),
as a function of energy $E_{0}$ by which the boson
energy is split. Strong dependence of $T_c$ on disorder
is a combined effect of the density of states, the
fluctuating interactions and the changes in
concentration of carriers.

%%%%%%%%%%%%%%%%%%%   Fig 5   %%%%%%%%%%%%%%%%%%%%%%%%%%
%%%%%%%%%%%%%%%%%%%%%%%%%%%%%%%%%%%%%%%%%%%%%%%%%%%%%%%%

%%%%%%%%%%%%%%%%%%%   Fig 6   %%%%%%%%%%%%%%%%%%%%%%%%%%
\begin{figure}
\centerline{\epsfxsize=8cm \epsfbox{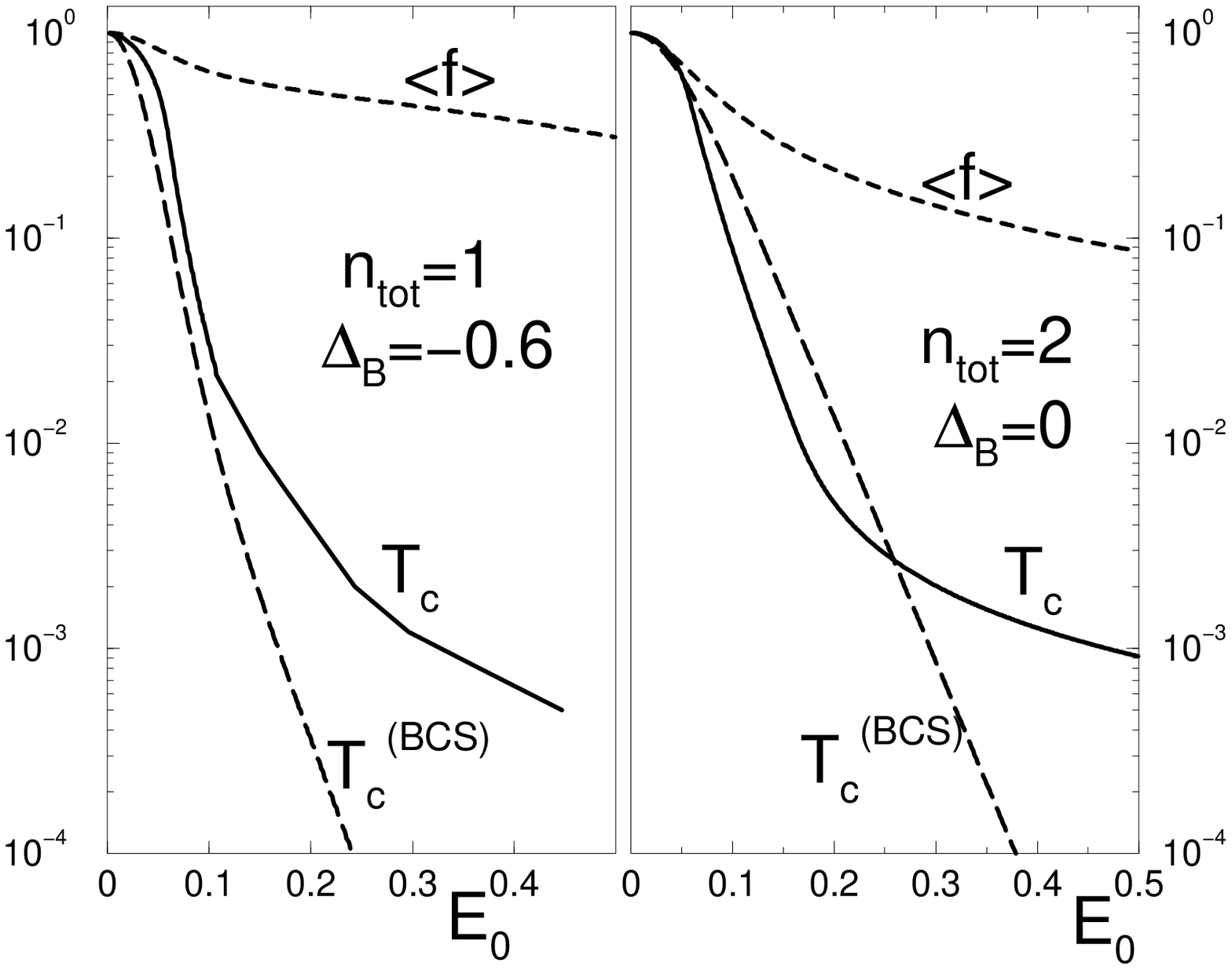}}
\vspace{3mm}
\caption{Normalized critical temperature $T_{c}/T_{c}
(E_{0}=0)$  and the normalized pairing potential  
$<f_{l}>/<f_{l}(E_{0}=0)>$ versus energy $E_{0}$. 
$T^{(BCS)}$ shows the BCS-like relation between
critical temperature and pairing potential.
Left panel refers to $n_{tot}=1$, $\Delta_{B}=-0.6$ 
discussed above and the right panel corresponds to 
the symmetric case of the BF model $\Delta_{B}=0$, 
$n_{tot}=2$ (with the half filled boson and fermion 
subsystems).} 
\label{fig6}
\end{figure}
%%%%%%%%%%%%%%%%%%%%%%%%%%%%%%%%%%%%%%%%%%%%%%%%%%%%%%%

To estimate what influence comes only from the 
renormalization of the effective pairing we plot
in figure \ref{fig6} the normalized transition 
temperature denoted by $T_c$ and the normalized 
averaged $<f_{l}>$ for the parameters given above 
(left panel), and for a fully symmetric case of 
the BF model (right panel). The transition 
temperature $T_c^{(BCS)}$ is the BCS-type estimate 
of the effect of changes in the effective pairing 
due to disorder
\begin{eqnarray} 
T_c(E_0) \propto \exp{ \left( \frac{-1}
{g(\varepsilon_F)<f_l(E_0)>} \right) }.
\end{eqnarray}
A general trend observed in figure \ref{fig6} is that 
the average effective interaction $<f_l(E_0)>$ decreases 
with increasing disorder, even though the bar fermion-boson 
interaction $v$ remains constant. This decreasing pairing 
interaction is the only factor responsible for a behavior 
of $T_c$ versus $E_{0}$ in the right panel. We notice 
absence of the BCS-like exponential scaling which is
due to unconventional pairing in the BF model.
 
In the left panel, corresponding to the above studied
case $\Delta_{B}=-0.6$, $n_{tot}=1$, we notice a larger
discrepancy between the pairing amplitude and $T_{c}$.
With an increase of $E_{0}$ the transition temperature 
is much strongly reduced than in the symmetric case.
This effect has to be assigned to redistributions of
particle occupancies. At large values of $E_{0}$ we
have practically only hard core boson particles in
the system, and they cannot induce superconductivity
among fermions whose fraction becomes very small. 

In the previous study \cite{Robaszkiewicz-01}
authors have used the same bimodal distribution 
of random boson energies. They have found a strong
reduction of $T_{c}$ near $E_{0} \sim 2v$ which 
agrees well with our data shown in figure 6.
Moreover, the authors have reported that disorder 
affects the ratio $\Delta_{sc}(T=0)/k_{B}T_{c}$ 
which changes from $4.2$ (for a clean system
\cite{Robaszkiewicz-87,Ranninger-95}) to the 
standard BCS result $3.52$ at large $E_{0}$. 
Simple explanation of this effect can be offered. 
The boson energy (which is split by $2E_{0}$) is 
for sufficiently large $E_{0}$ partly in the LP 
limit (for $E_{i}=-E_{0}$) and partly in the BCS 
limit (if $E_{i}=+E_{0}$). The second one 
contributes with the standard BCS value
if $|E_{0}|$ is large enough (see e.g.\ Fig.~9
in Ref.\ \cite{Robaszkiewicz-87}).

\section{Conclusion}

The randomness of the site energies of both, 
fermions and bosons, has a strong effect on 
superconducting phase of the BF model. Weak disorder 
in the fermion subsystem affects the superconducting 
transition temperature mainly via rescaling the low 
energy states which are involved in the the formation 
of the Cooper pairs. Therefore $T_c$ roughly follows 
the density of states at the Fermi energy. For 
sufficiently large disorder $\varepsilon_{0}$ there 
appears some critical concentration $c$ at which 
$T_{c}$ may eventually drop to zero.

Disorder in the boson subsystem has a much more fine 
influence on superconductivity. Randomness of boson 
energies is transformed directly into randomness of 
the pairing strength. Effectively physics of the 
disordered BF model becomes similar to that of the 
random negative $U$ Hubbard model \cite{Litak_00}.
Even the relatively small fluctuations of the boson 
energies show up their strong detrimental effects 
on superconductivity. 

In a simple minded picture one can envision this 
situation as a random change between various regimes 
of superconductivity. Depending on a value of $E_l$ 
the boson energy $\Delta_{B}+E_l$ can be either far
below the Fermi energy (the LP limit), or far above 
the Fermi energy (the BCS limit). Each of such random 
configurations contributes with a different strength 
of superconducting correlations. On average, the 
superconducting transition temperature $T_c$ strongly 
diminishes and practically disappears if the amplitude 
of the randomly fluctuating boson energies $|E_{0}|$ 
is large enough. 

In summary, our calculations show that disorder 
strongly affects the $s$-wave superconducting phase 
of the BF model. This apparent contradiction with 
Anderson theorem can be understood because of a 
change of the effective pairing interaction induced
by disorder, and this effect is contrary to the 
Anderson's main assumption \cite{Anderson-59}. 

To compare our results with experimental data on high 
temperature superconductors one has to consider the 
$d$-wave superconducting order parameter. This type of 
a symmetry arises in a natural way according to the 
recent derivation of the BF model \cite{plaquette-96}.
Effect of disorder on such anisotropic superconducting
phase of the BF model is outside the scope of the present 
paper and will be discussed elsewhere.

\acknowledgements{
We would like to thank Julius Ranninger for helpful discussions. 
This work has been partly supported by the Polish 
State Committee for Scientific Research under project 
No.\ 2P03B 106 18. T.D.\ kindly acknowledges 
hospitality of the Joseph Fourier University and 
the Centre de Recherches sur les Tr\`es Basses 
Temperatures in Grenoble, where part of this 
study has been done.}

\end{multicols}
\end{document}